%% file: Disomega.tex
\documentstyle[epsfig,subfig,amsmath,array]{elsarticle}
\newcommand{\newc}{\newcommand}
\newc{\lra}{\leftrightarrow}
\newc{\beq}{\begin{equation}}
\newc{\eeq}{\end{equation}}
\newc{\barr}{\begin{eqnarray}}
\newc{\earr}{\end{eqnarray}}

\input{begin.tex}

\begin{document}
\def\vbf{\mbox{\boldmath $\upsilon$}}
\def\barr{\begin{eqnarray}}
\def\earr{\end{eqnarray}}
\def\g{\gamma}
\newcommand{\dphi}{\delta \phi}
\newcommand{\bupsilon}{\mbox{\boldmath \upsilon}}
\newcommand{\bfup}{\mbox{\boldmath \upsilon}}
\newcommand{\at}{\tilde{\alpha}}
\newcommand{\pt}{\tilde{p}}
\newcommand{\Ut}{\tilde{U}}
\newcommand{\rhb}{\bar{\rho}}
\newcommand{\pb}{\bar{p}}
\newcommand{\pbb}{\bar{\rm p}}
\newcommand{\kt}{\tilde{k}}
\newcommand{\wt}{\tilde{w}}

\date{\today}
\title {Axionic dark matter signatures in  various halo models}
%
%
%
%
%
\author{, J.D. Vergados$^{1,2}$ and Y. Semertzidis$^{1}$}
%
%
\address{
1 KAIST University, Daejeon, Republic of Korea and \\
Center for Axion and Precision Physics Research, IBS, Daejeon 305-701, Republic of Korea
}
\address{ARC Centre of Excellence in Particle Physics at the Terascale and Centre for the Subatomic Structure of Matter (CSSM), University of Adelaide, Adelaide SA 5005, Australia\footnote{Physics Department, Univesity of Ioannina. e-mail:vergados@uoi.gr}}
\begin{frontmatter}
\begin{abstract}
In the present work we study possible time signatures  Axion Dark Matter searches  employing  resonant cavities for various halo models. We study in particular the time dependence  of the resonance width (modulation) and possible asymmetries  in directional experiments.
\end{abstract}

\end{frontmatter}
\section{Introduction}
The axion has been proposed a long time ago as a solution to the strong CP problem \cite{PecQui77} resulting to a pseudo Goldstone Boson \cite{SWeinberg78,Wilczek78,PWW83,AbSik83,DineFisc83}, but it has also been recognized as a prominent  dark matter candidate \cite{PriSecSad88}.
 In fact, realizing an idea proposed a long time ago by Sikivie \cite{Sikivie83}, various experiments such as ADMX and ADMX-HF collaborations 
\cite{Stern14,MultIBSExp}, \cite{ExpSetUp11b},\cite{ADMX10} are now planned to search for them. In addition, the newly established center for axion and physics research (CAPP)  has started an ambitious axion dark matter research program \cite{CAPP}, using SQUID and HFET technologies \cite{ExpSetUp11a}.
  The allowed parameter space  \cite{MultIBSTh}, containing information for all the axion like particles, defines a  region for invisible axions, which can be dark matter candidates.
	
In the present work we will assume that the axion is non relativistic with mass in $\mu$eV-meV scale, moving with an average velocity which is $\approx0.8\times10^{-3}$c.  It is expected to be observed in resonance cavities with a  width that depends  on the axion mean square velocity in the local frame. The latter depends on the assumed halo model. We will expand and improve our previous work \cite{SemerVer15,SemerVerProc} by including   various popular halo models.  We will study the   time variation of the width due to the motion of the Earth around the sun and possible asymmetries with regard to the sun's direction of motion as it goes around the center of the galaxy. Due to the rotation of the Earth around its axis these asymmetries in the width of the resonance will manifest themselves in their diurnal variation. These important signatures may become more needed, if indeed the predicted axion to photon coupling becomes weaker as  some recent models predict \cite{Ahn14}.
\section {Brief summary of the formalism}
The photon axion interaction is dictated by the Lagrangian:
\beq
{\cal L}_{a \gamma\gamma}=g_{a\gamma\gamma}a {\bf E}\cdot{\bf B},\, g_{a\gamma\gamma}=\frac{\alpha g_{\gamma}}{ \pi f_a},
\eeq
where  ${\bf E}$ and ${\bf B}$ are the electric and magnetic fields, $g_{\gamma}$ a model dependent constant of order one 
\cite{Stern14},\cite{HKNS14}\cite{JEKim98} and $f_a$ the axion decay constant. Axion dark matter detectors \cite{HKNS14} employ an external magnetic field, ${\bf B}\rightarrow {\bf B}_0$ in the previous equation, in which case one of the photons is replaced by a virtual photon, while the other maintains the energy of the axion, which is its mass plus a small fraction of kinetic energy. \\
The power produced, see e.g.  \cite{Stern14}, is given by:
\beq
P_{mnp}=g_{a\gamma\gamma}^2\frac{\rho_a}{m_a} B_0^2 V C_{mnp} Q_L
\label{Eq:Pmnp}
\eeq
$Q_L$ is the loaded quality factor of the cavity. Here we have assumed $Q_L$ is smaller than the axion width $Q_a$, see below.  More generally, $Q_L$ should be substituted by min ($Q_L$, $Q_a$).
This power depends on the axion density and is pretty much independent of the velocity distribution.
	
The axion power spectrum, which is of great interest to experiments, is written as a Breit-Wigner shape \cite{HKNS14}, \cite{KMWM85}:
\beq
\left |{\cal A}(\omega) \right |^2=\frac{ \rho_D}{m_a^2}\frac{\Gamma}{(\omega-\omega_a)^2+(\Gamma/2)^2},\Gamma=\frac{\omega_a}{Q_a}
\eeq
 Since  in the
axion DM search case the cavity detectors have reached such a very high energy resolution \cite{Duffy05,Duffy06}, one should try to accurately evaluate the width of the expected power spectra in various theoretical models.
\section{Evaluation of the width in the local frame}
We will derive the expression for the width assuming in the galactic frame a Maxwell Boltzmann axion velocity distribution:
\beq
f(\upsilon)=\frac{1}{\pi\sqrt{\pi}}\frac{1}{\upsilon_0^3}e^{-\frac{\upsilon^2}{\upsilon^2_0}}
\eeq
We will use the relation
\beq 
\omega=m_a \left (1+\frac{1}{2}\upsilon^2 \right )
\eeq
or
$$\upsilon=\sqrt{\frac{2(\omega-m)}{m}}, \upsilon d \upsilon=\frac{d \omega}{m}$$
In the galactic frame the number of axions in the with frequency between $\omega$ and $\omega+d \omega$ in a solid angle $d \Omega$
is
\beq
d N_a=\frac{\rho_a}{m_a}\frac{1}{\pi\sqrt{\pi}}\frac{1}{\upsilon_0^3}e^{-2\frac{\omega-m}{m\upsilon^2_0}}\sqrt{\frac{2(\omega-m)}{m}}\frac{d \omega}{m}d \Omega
\eeq
Introducing the variable $x=2\frac{\omega-m}{m\upsilon^2_0}$  we find:
\beq
d N_a=\frac{\rho_a}{m_a}g(x)\frac{d \Omega}{4 \pi},\,g(x)=\frac{2}{\sqrt{\pi}} \sqrt{x}e^{-x}
\label{Eq:local}
\eeq
The maximum of the distribution occurs at $x=1/2$. The width at half maximum is $\delta x=x_2-x_1$ with $x_i$  the roots of the equation 
$$g(x)=\frac{1}{2}\left . g(x) \right |_{x=1/2}\Rightarrow \sqrt{x}e^{-x}=\frac{1}{2 \sqrt{2}}e^{-\frac{1}{2}}$$
We thus find $\delta x=1.8$. Thus the width at half maximum in frequency space is:
\beq
\delta  \omega=m \frac{1}{2} \upsilon^2_0 \delta x \Rightarrow \frac{1}{Q_a}=\frac{1}{2}\upsilon^2_0 \delta x= \delta x \frac{1}{3}\prec\upsilon^2\succ _g
\eeq
where the last quantity is the average of the square of the axion velocity in the galactic frame.

Our next task is to transform the velocity distribution from the
galactic to the local frame. The needed equation, see e.g.
\cite{Vergados12}, is:
   \beq
{\bf y} \rightarrow {\bf y}+{\hat\upsilon}_s+\delta \left
(\sin{\alpha}{\hat x}-\cos{\alpha}\cos{\gamma}{\hat
y}+\cos{\alpha}\sin{\gamma} {\hat \upsilon}_s\right ) ,\quad
y=\frac{\upsilon}{\upsilon_0} \label{Eq:vlocal} \eeq with
$\gamma\approx \pi/6$, $ {\hat \upsilon}_s$ a unit vector in the
Sun's direction of motion, $\hat{x}$  a unit vector radially out
of the galaxy in our position and  $\hat{y}={\hat
\upsilon}_s\times \hat{x}$. The last term in the first expression
of Eq. (\ref{Eq:vlocal}) corresponds to the motion of the Earth
around the Sun with $\delta$ being the ratio of the modulus of the
Earth's velocity around the Sun divided by the Sun's velocity
around the center of the Galaxy, i.e.  $\upsilon_0\approx 220$km/s
and $\delta\approx0.135$. The above formula assumes that the
motion  of both the Sun around the Galaxy and of the Earth around
the Sun are uniformly circular. The exact orbits are, of course,
more complicated  but such deviations are not
expected to significantly modify our results. In Eq.
(\ref{Eq:vlocal}) $\alpha$ is the phase of the Earth ($\alpha=0$
around the beginning of June)\footnote{One could, of course, make the time
dependence of the rates due to the motion of the Earth more
explicit by writing $\alpha \approx(6/5)\pi\left (2 (t/T)-1 \right
)$, where $t/T$ is the fraction of the year.}.

\begin{figure}[!ht]
\begin{center}
\subfloat[]
{
\rotatebox{90}{\hspace{-0.1cm} {$g(x,\xi)\rightarrow$}}
\includegraphics[width=0.45\textwidth, height=0.3\textwidth]{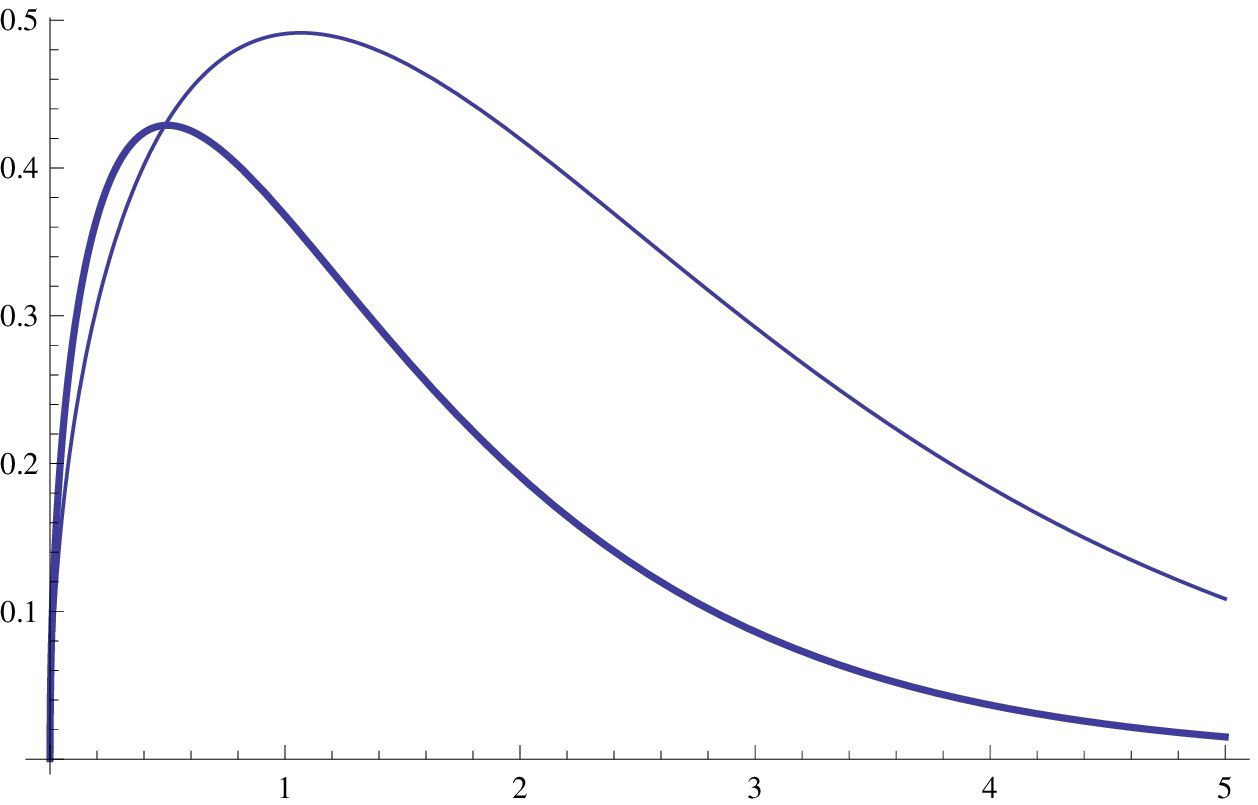}
}
\subfloat[]
{
\rotatebox{90}{\hspace{-0.1cm} {$g(x,\xi) \rightarrow$}}
\includegraphics[width=0.45\textwidth, height=0.3\textwidth]{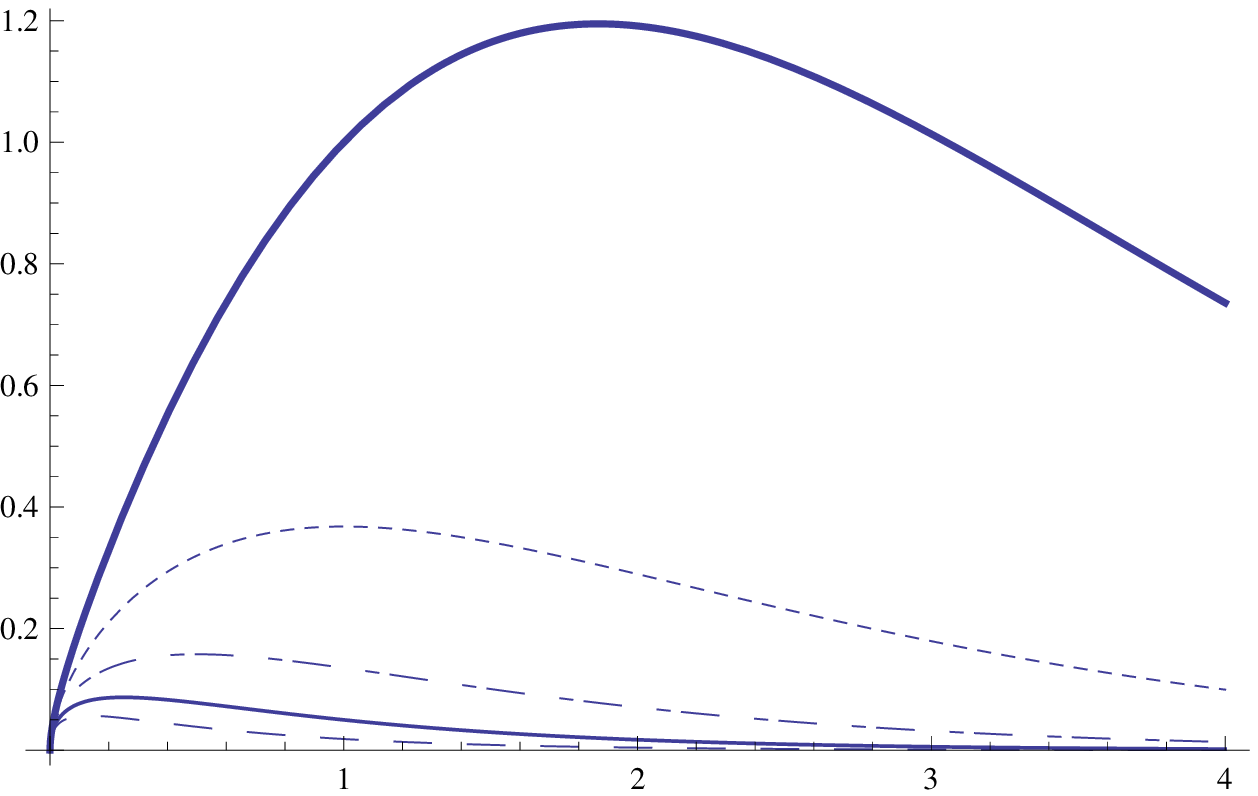}
}\\
{\hspace{-2.0cm} {$x\rightarrow$ }}\\
 \caption{The  function $g(x,\xi)$, $x=2\frac{\omega-m}{m\upsilon^2_0}$, which essentially gives the normalized frequency distribution, is exhibited. In (a) we show the distribution in the galactic frame (thick solid line) and its angle average in the local frame (solid line). In (b) we exhibit the angular dependence of the distribution in the local frame for some typical values of $\xi$, $xi=\cos{\theta}$, $\theta$ being the polar angle of the velocity  with respect to the sun's direction of motion. The motion of the Earth is ignored. From top to bottom $\xi=-1,-1/2,0,1/2,1$. The asymmetry is obvious.}
 \label{fig:disomega}
  \end{center}
  \end{figure}

 In the local frame , ignoring the motion of the Earth, we make in the velocity distribution the substitution:
$$\upsilon^2\rightarrow\upsilon^2+\upsilon_0^2+2 \upsilon \upsilon_0 \xi,\,\xi=\hat{\upsilon}.\hat{\upsilon}_z$$
where $\upsilon_0$ is the speed of the sun around the center of the galaxy and $\hat{\upsilon}_z$, is a  unit vector in the sun's direction of motion.
Thus Eq. (\ref{Eq:local})
\beq
d N_a=\frac{\rho_a}{m_a}g(x,\xi)\frac{d \Omega }{4 \pi},\,g(x,\xi)=\frac{2}{\sqrt{\pi}} \sqrt{x}e^{-\left (1+x+2 \xi \sqrt{x}\right )}
\label{Eq:nonlocal}
\eeq
The function$g(x,\xi) $
is exhibited in Fig. \ref{fig:disomega}.

\begin{figure}[!ht]
\begin{center}
\subfloat[]
{
\rotatebox{90}{\hspace{-0.0cm} {$(\delta x(\mbox{dir})/\delta x(\mbox{non dir}) \rightarrow$}}
\includegraphics[width=0.45\textwidth, height=0.3\textwidth]{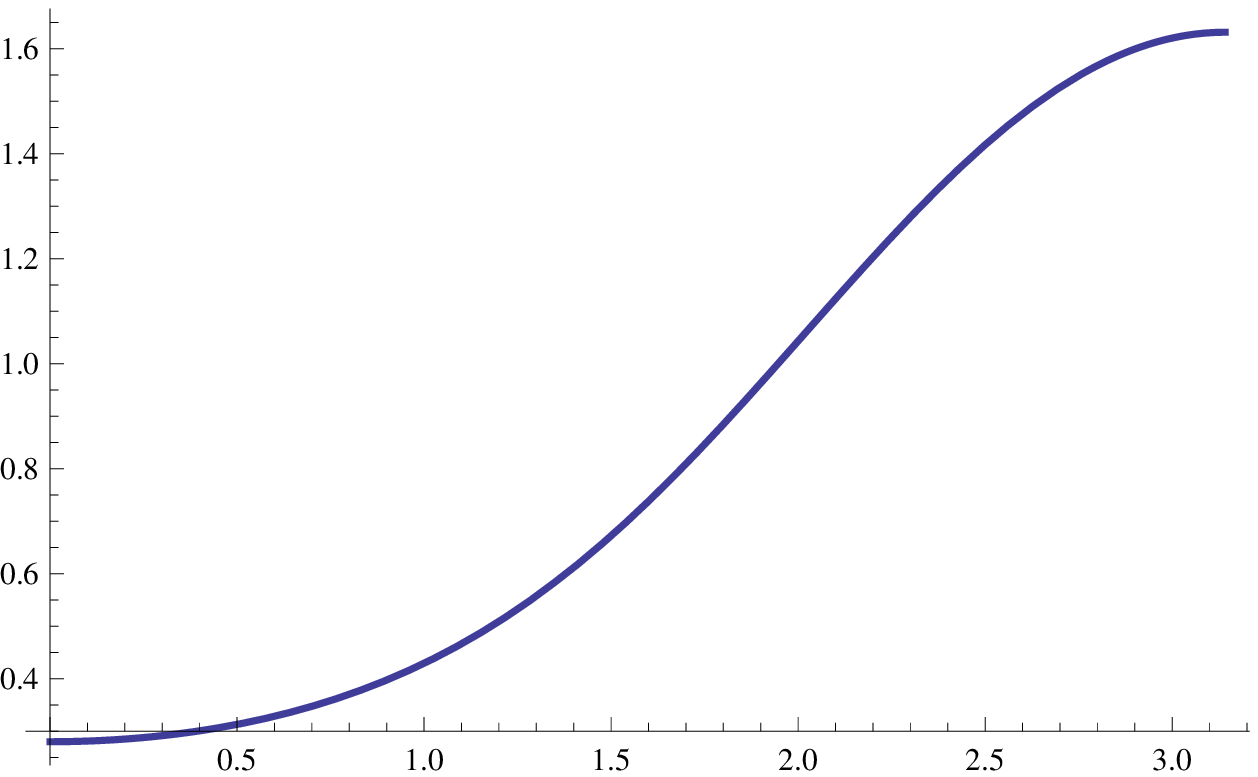}
}
\subfloat[]
{
\rotatebox{90}{\hspace{-0.0cm} {$(\delta x(\mbox{dir})/\delta x(\mbox{non dir}) \rightarrow$}}
\includegraphics[width=0.45\textwidth, height=0.3\textwidth]{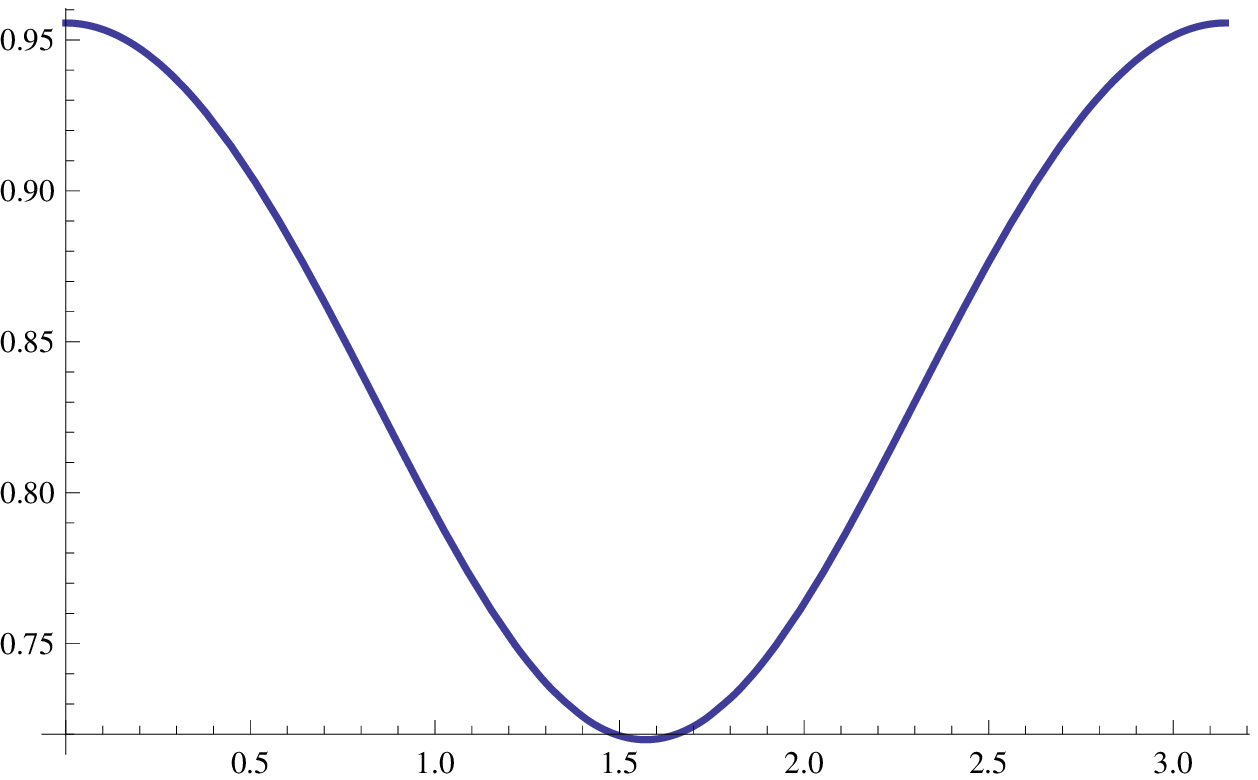}
}\\
{\hspace{-2.0cm} {$\Theta\rightarrow$ radians}}\\
 \caption{The  width $\delta x$ expected in directional experiments, relative to that of the standard (non directional) experiments,  as a function of the polar angle with sense of direction known (a) and  with sense of direction not known (b). 
 \label{fig:DirAxionWidth}}
  \end{center}
  \end{figure}

$ g(x,\xi) $ has extrema at 
$$x=\left\{\frac{1}{2} \left(\xi ^2-\sqrt{\xi ^4+2 \xi
   ^2}+1\right),\frac{1}{2} \left(\xi ^2+\sqrt{\xi ^4+2
   \xi ^2}+1\right)\right\}
	$$
	The first of these yields a maximum for all $\xi$, while the second is the location of a maximum only for $-1\leq \xi \leq \frac{1}{4}\sqrt{19 - 3 \sqrt{33}}=0.33 $.\\
	Clearly the width at half maximum is a function of $\xi$. Its value is obtained numerically and it is exhibited in Fig. \ref{fig:DirAxionWidth}a,b, corresponding to whether the sense of direction of the axion can be determined or not.  The width clearly has a maximum opposite to the sun's direction of motion (a), if the sense of direction of the axion  can be determined, something perhaps not very realistic. It exhibits  simply a minimum in the plane perpendicular to the sun's velocity, in the most realistic scenario that  the sense of direction cannot be determined (b).
	
\begin{figure}[!ht]
\begin{center}
\rotatebox{90}{\hspace{-0.1cm} {$\delta x (\alpha)/\delta x (\pi/2) \rightarrow$}}
\includegraphics[width=0.9\textwidth, height=0.6\textwidth]{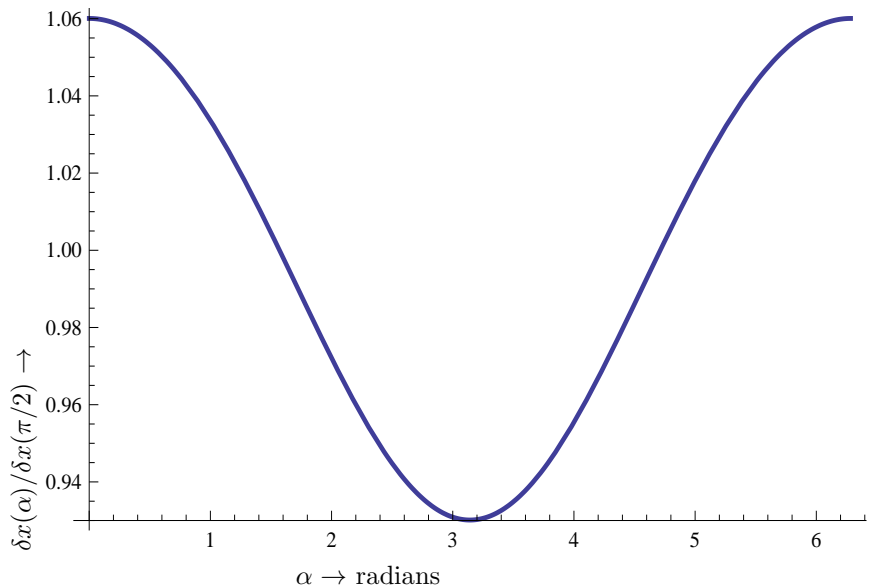}\\
{\hspace{-2.0cm} {$\alpha\rightarrow$ radians}}\\
 \caption{The  modulated width $\delta x$, relative to its time average, as a function of the phase of the Earth $\alpha$ ($\alpha=0$ on June 3nd).}
 \label{fig:ModWidth}
  \end{center}
  \end{figure}

	We will next consider the effect of the Earth's motion in the standard, non directional, experiments. The  azymouthal angle  ($\phi$) dependence,  averages out to zero. Thus the distribution takes the form:
	\beq
d N_a=\frac{\rho_a}{m_a}h(x,\xi,\cos{\alpha})\frac{d \Omega }{4 \pi},\,h(x,\xi,\cos{\alpha})=\frac{2}{\sqrt{\pi}} \sqrt{x}e^{-\left (1+x+(2 +\delta \cos{\alpha})\xi \sqrt{x}\right )}
\label{Eq:modulation}
\eeq
$$g(x,\cos{\alpha})=\int h(x,\xi,\cos{\alpha})\frac{d \Omega }{4 \pi}=\frac{2 e^{-x-1} \sinh \left(\sqrt{x} (\delta  \cos (\alpha
   )+2)\right)}{\sqrt{\pi } \sqrt{x} (\delta  \cos (\alpha )+2)}
	$$
The maximum of the distribution $g(x,\cos{\alpha}$ occurs at the root $r_1(\cos{\alpha})$  of the equation :
$$ \coth \left(\sqrt{x} (\delta  \cos (\alpha
   )+2)\right)-\frac{\sqrt{x}}{\delta  \cos (\alpha )+2}=0$$
	which is obtained graphically. Then  $\delta x(\cos{\alpha})=x_2(\cos{\alpha})-x_1(\cos{\alpha})$ where $x_1(\cos{\alpha})$ and $x_2(\cos{\alpha})$ are the roots of the equation:
	$$ g(x,\cos{\alpha})=(1/2)g(r_1(\cos{\alpha}),\cos{\alpha})$$
	where
$$ g(x,\cos{\alpha})=\sqrt{x}e^{-\left (1+x+(2 +\delta \cos{\alpha})\xi \sqrt{x}\right )}$$
The obtained results are exhibited in Fig. \ref{fig:ModWidth}.
	\section{More complicated velocity distributions}
	In the context of dark matter other velocity distributions have been considered, e.g. completely phase-mixed DM, dubbed ``debris flow''
(Kuhlen et al.~\cite{Spergel12}) and caustic rings  (Sikivie~\cite{SIKIVI1},\cite{SIKIVI2}\cite{Sikivie08}, Vergados~\cite{Verg01}.
\subsection{The case of debris flows}
\begin{figure}[!ht]
\begin{center}
\rotatebox{90}{\hspace{-0.1cm} {$\delta x (\alpha)/\delta x (\pi/2) \rightarrow$}}
\includegraphics[width=0.9\textwidth, height=0.6\textwidth]{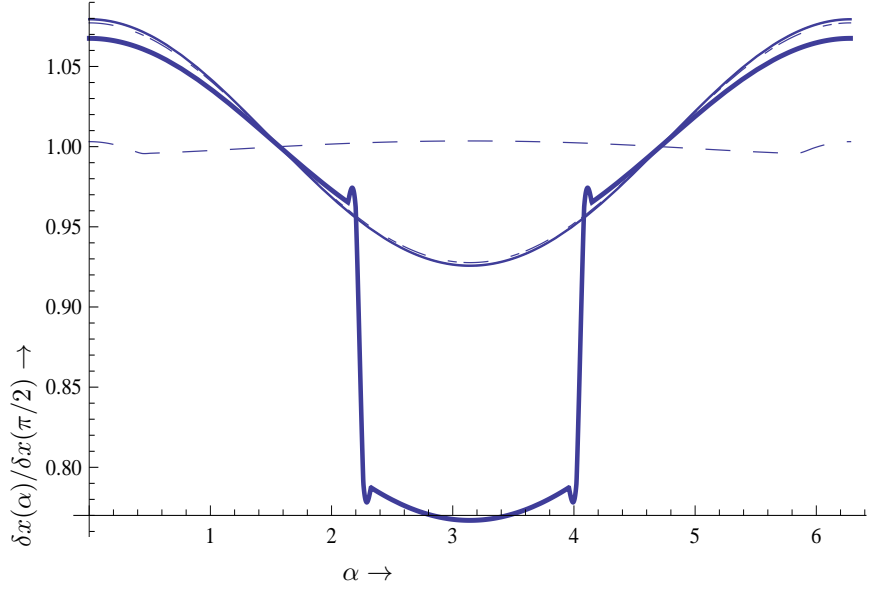}\\
{\hspace{-2.0cm} {$\alpha \rightarrow$ }}\\
 \caption{The  modulated width, relative to its time average, as a function of the phase of the Earth  for various values of $\epsilon$ .  $\epsilon=0.220,\,0.222,\,0.241,\,0.323\,0.508$ is associated with thick solid, dotted, dot-dashed, fine solid and dashed lines respectively. The dotted and the fine solid lines cannot be resolved.}
 \label{fig:ModDFWidth}
  \end{center}
  \end{figure}

\begin{figure}[!ht]
\begin{center}
\subfloat[]
{
\rotatebox{90}{\hspace{-0.1cm} {${\delta x({\mbox{{dir}}})}/ {\delta x({\mbox{{non dir}}})}  \rightarrow$}}
\includegraphics[width=0.45\textwidth, height=0.3\textwidth]{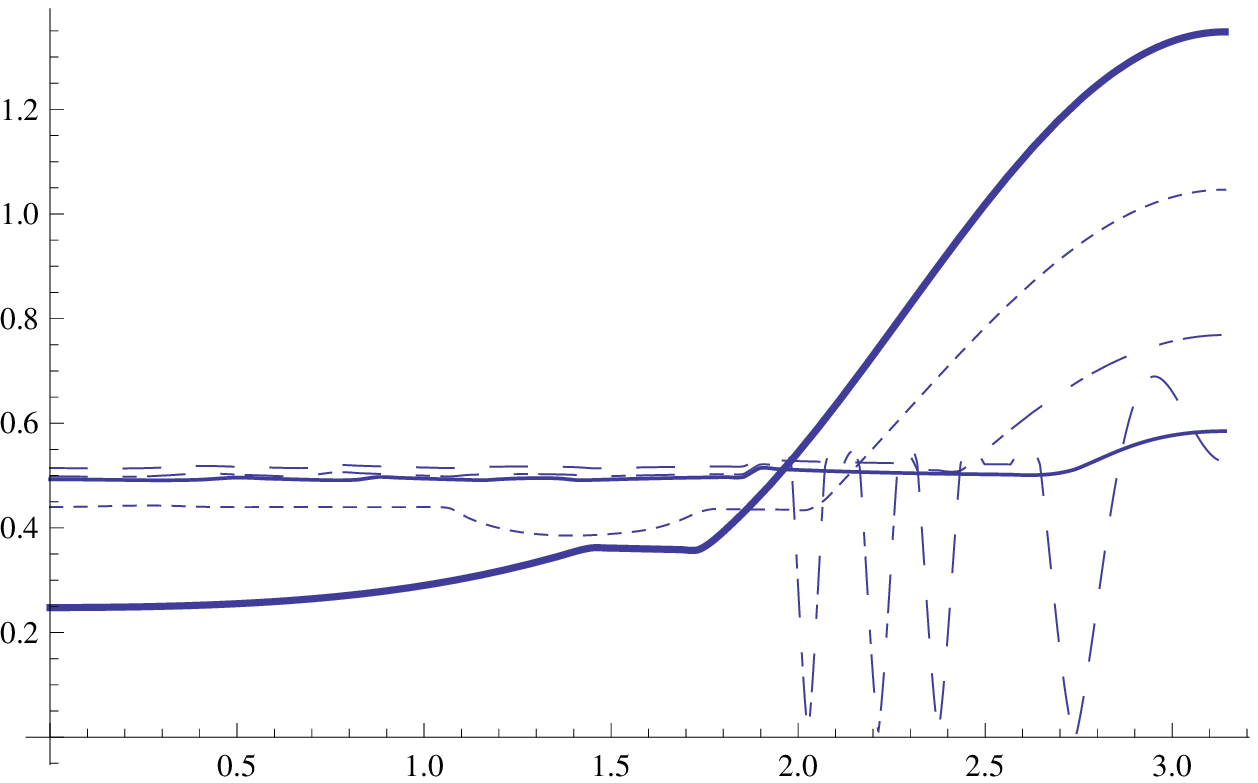}
}
\subfloat[]
{
\rotatebox{90}{\hspace{-0.1cm} {${\delta x({\mbox{{dir}}})}/ {\delta x({\mbox{{non dir}}})}  \rightarrow$}}
\includegraphics[width=0.45\textwidth, height=0.3\textwidth]{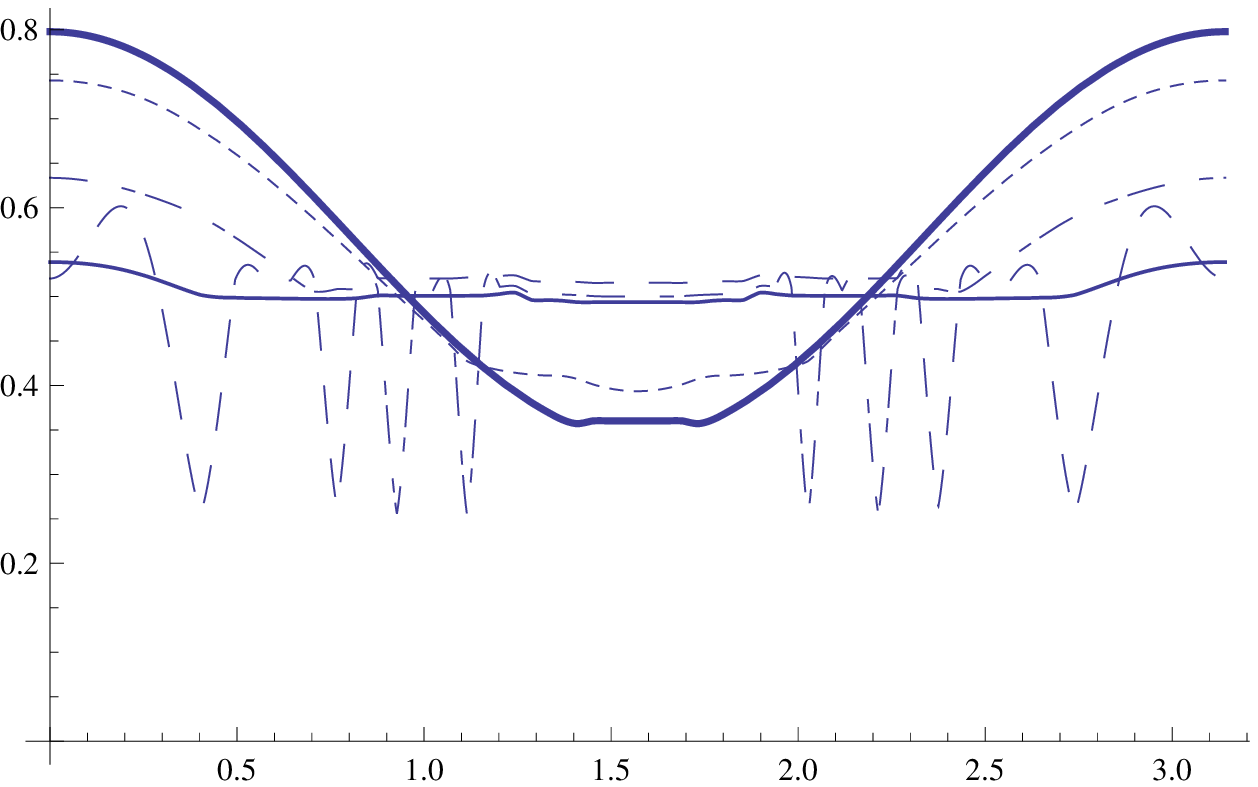}
}\\
{\hspace{-2.0cm} {$\Theta \rightarrow$ }}\\
 \caption{The  width $\delta x$ expected in directional experiments, relative to that expected in standard experiments, as a function of the angle between the  direction of observation and the sun's direction of motion, in the presence of debris flows. The exhibited results correspond to the two cases: The  sense of direction is known (a) or is not determined (b).  Otherwise the notation is the same as in Fig.    \ref{fig:ModDFWidth}.
 We see that the effect of the debris flows is to decrease  $\delta x$ in all directions.
 \label{fig:DirDFWidth}}
  \end{center}
  \end{figure}

 In the case of debris flows one assumes a dark matter distribution which is  of the form:
\beq
f(\upsilon)=(1-\epsilon(\zeta))f_{MB}(\upsilon)+\epsilon(\zeta) f_{db}(\upsilon)
\eeq
with
\beq
f_{db}(\upsilon)=\left \{ \begin{array}{cc}\frac{1}{2\upsilon_f \upsilon_0}\upsilon,&\upsilon_f-\upsilon_0\leq\upsilon\leq\upsilon_f+\upsilon_0\\0&\mbox{otherwise} \end{array} \right .
\eeq
where 
\beq
\epsilon(\zeta)=0.22 + 0.34 \left (\mbox{erf}\left (\zeta \frac{220}{185}-\frac{465}{185}\right ) + 1\right )
\label{Eq:epsilonmix}
\eeq
\begin{figure}[!ht]
\begin{center}
\rotatebox{90}{\hspace{-0.1cm} {$\epsilon(\zeta)  \rightarrow$}}
\includegraphics[width=0.9\textwidth, height=0.6\textwidth]{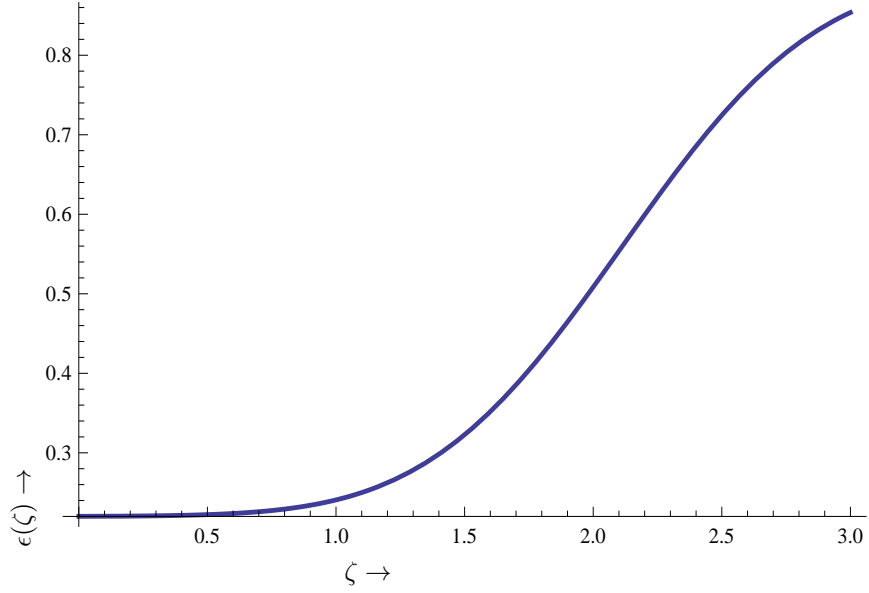}\\
{\hspace{-2.0cm} {$\zeta \rightarrow$ }}\\
 \caption{The  amplitude $\epsilon(\zeta)$ entering the debris flow component of dark matter. Of particular interest is the region of $\zeta$ between 1.5 and 2.}
 \label{fig:dfepsilon}
  \end{center}
  \end{figure}
	The introduction of debris flows has two effects:
	\begin{itemize}
	\item It shifts the maximum of the distribution, without changing the location of the maximum $r_1$.
	\beq
	g(x)\rightarrow (1-\epsilon(\zeta))g(x)+\epsilon(\zeta) g_{df}(x)
	\eeq
	with
	\beq
g_{df}(x)=\left \{ \begin{array}{cc}\frac{1}{4(1+\zeta)},&\zeta^2 \leq x\leq (2+\zeta)^2\\0&\mbox{otherwise} \end{array} \right .
\eeq
	\item The width at half maximum is affected:
	\beq
	g(x)=\frac{1}{2}g(r_1)\Leftrightarrow (1-\epsilon(\zeta))g(x)+\epsilon(\zeta) g_{df}(x)
=(1-\epsilon(\zeta))g(r_1)+\epsilon(\zeta) g_{df}(r_1)
	\eeq
	\end{itemize}
	\begin{figure}
\begin{center}
\subfloat[]
{
\rotatebox{90}{\hspace{0.0cm} $\epsilon(x)\rightarrow$}
\includegraphics[width=0.45\textwidth,height=0.3\textwidth]{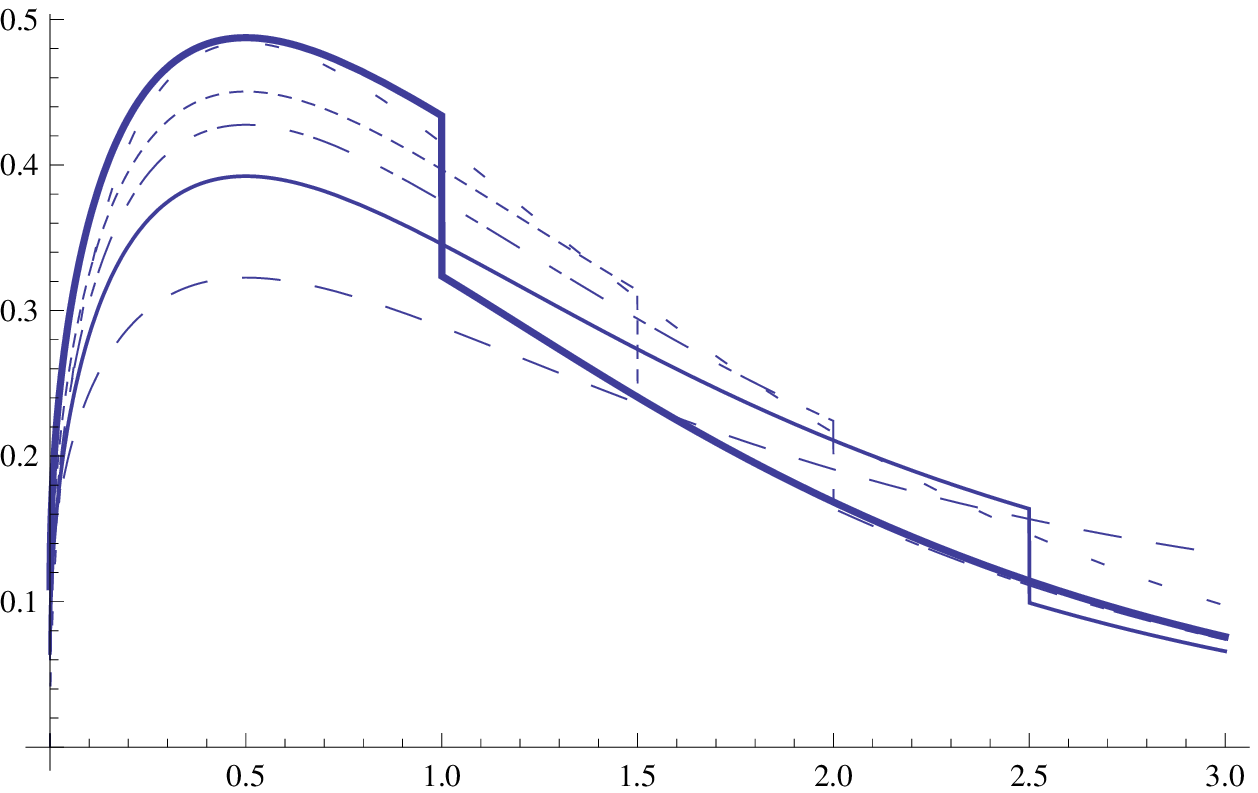}
}
\subfloat[]
{
\rotatebox{90}{\hspace{0.0cm} $\epsilon(x)\rightarrow$}
\includegraphics[width=0.45\textwidth,height=0.3\textwidth]{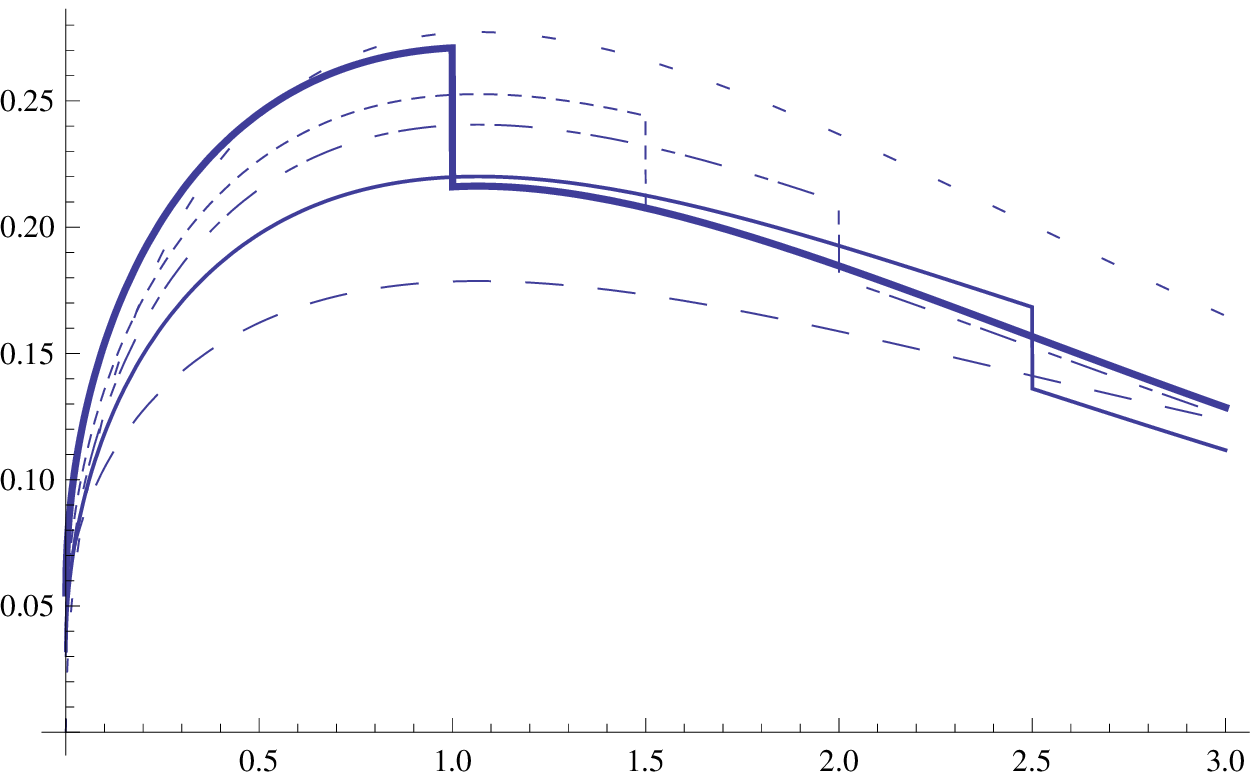}
}
\\
{\hspace{-2.0cm} $x=\rightarrow$}
\end{center}
\caption{ The  frequency distribution  function $g(x)\rightarrow (1-\epsilon)g(x)+\epsilon g_{df}(x)$ with respect to the galaxy (a) and in the local frame (b). The top line corresponds to $g(x)$ only. Otherwise  from top to bottom the lines correspond to $\epsilon=0.220,0.222,0.241,0.323$ and $0.508$.
 \label{fig:epsilon}}
\end{figure}
We thus find:
\begin{itemize}
\item In the galactic frame:\\
$$\delta x=(1.82,\, 1.81,\,1.97,\,\,2.11,\,2.40) \mbox{ for } \epsilon=(0.220,\,0.222,\,0.241,\,0.323,\,0.508 )$$ respectively
\item In the local frame:\\
$$\delta x=(3.06,\, 3.23,\,3.28,\,\,3.23,\,2.96) \mbox{ for } \epsilon=(0.220,\,0.222,\,0.241,\,0.323,\,0.508) $$
 respectively
\end{itemize}
The modulation of the width is exhibited in Fig. \ref{fig:ModDFWidth}.
The width expected in directional experiments is shown in Fig. \ref{fig:DirDFWidth}.
	\subsection{The case of caustic rings}
	Our study of the phase space structure of the Milky Way halo is motivated in large part
by the ongoing searches for dark matter on Earth, using axion, see e.g. \cite{ExpSetUp11a,ExpSetUp11b}, and WIMP detectors, see e.g. \cite{XENON100.11,DamaSasso13,EDELWEISS11,CDMS05,XMASS09}.
The signal in such detectors depends on the velocity distribution of dark matter in the solar
neighborhood. The caustic ring halo model predicts , see e.g. \cite{Sikivie08},  that most of the local dark matter is
in discrete flows and provides the velocity vectors and densities of the first forty flows at the Earth’s location in the Galaxy, which are essential when interpreting a
signal in a dark matter detector on Earth. In principle the halo model can be treated as phase-mixed DM leading to a linear combination of a M-B distribution and one appropriate to caustic rings, in a fashion  similar to that with debris flow  discussed above. We will, however, concentrate  here in caustic rings. 

\begin{figure}[!ht]
\begin{center}
\rotatebox{90}{\hspace{-0.1cm} {$(\delta x/(\delta x)_{av}) \rightarrow$}}
\includegraphics[width=0.9\textwidth, height=0.4\textwidth]{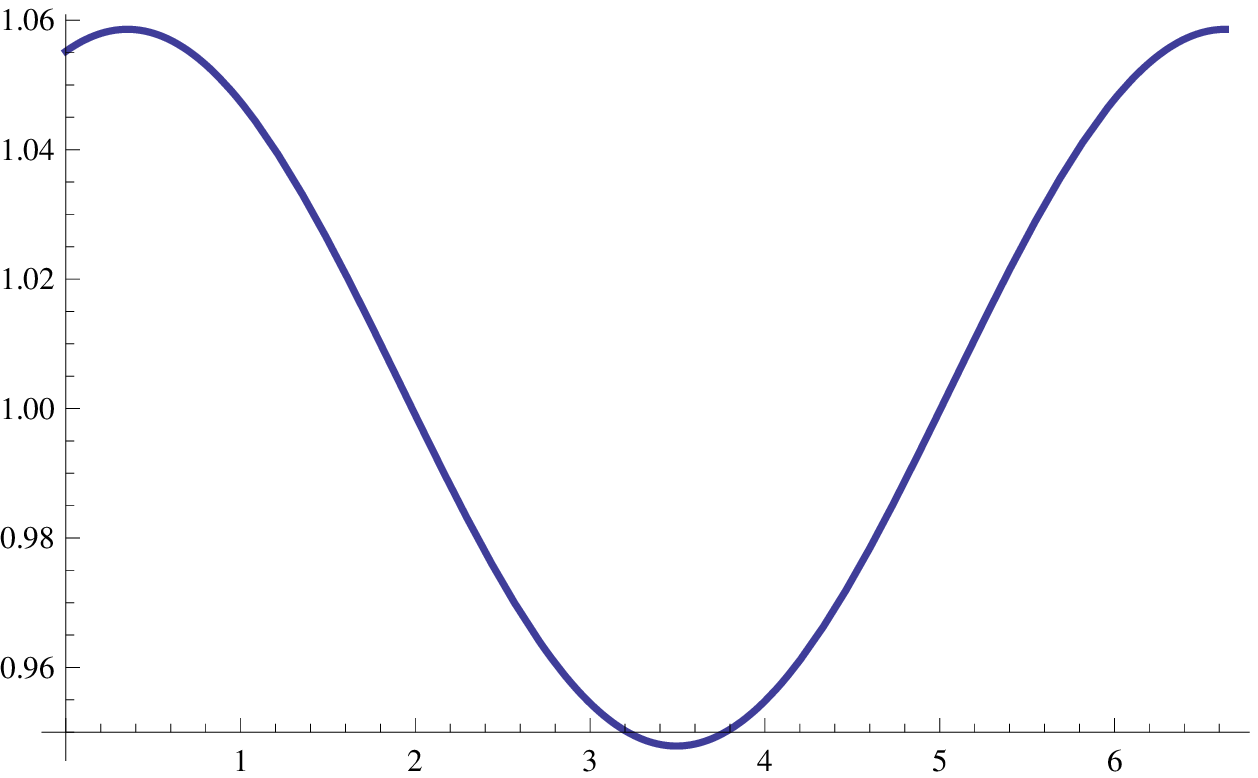}
\\
{\hspace{-2.0cm} {$\alpha \rightarrow$ }}
 \caption{The  width modulated width  $\delta x$ relative to the time averaged width expected in the case of caustic rings. Note that in this case the maximum does not occur at $\alpha=0$, but a bit later.}
 \label{fig:cModWidth}
  \end{center}
  \end{figure}

\begin{figure}[!ht]
\begin{center}
\subfloat[]
{
\rotatebox{90}{\hspace{-0.1cm} {$(\delta x(\mbox{dir})/\delta x(\mbox{non dir}) \rightarrow$}}
\includegraphics[width=0.9\textwidth, height=0.4\textwidth]{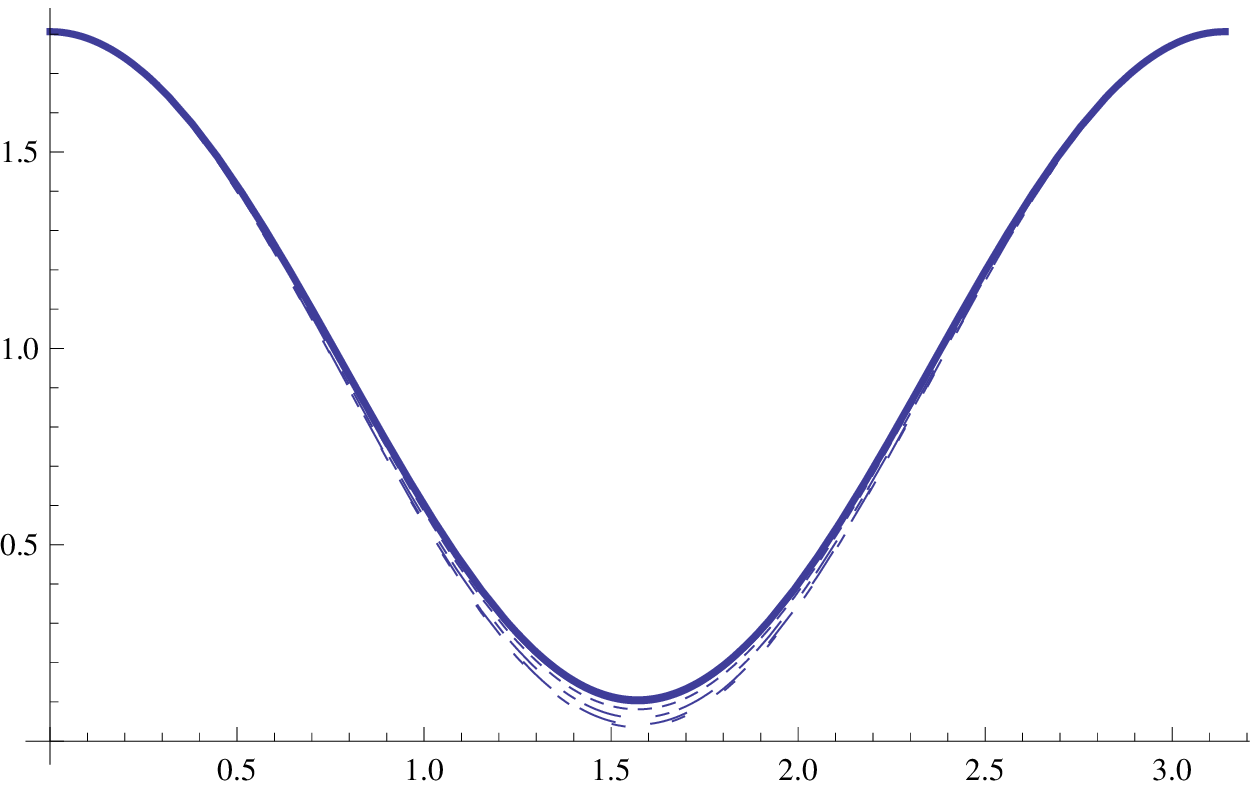}
}
\\
{\hspace{-2.0cm} {$\Theta \rightarrow$ }}\\
\subfloat[]
{
\rotatebox{90}{\hspace{-0.1cm} {$(\delta x(\mbox{dir})/\delta x(\mbox{non dir}) \rightarrow$}}
\includegraphics[width=0.9\textwidth, height=0.4\textwidth]{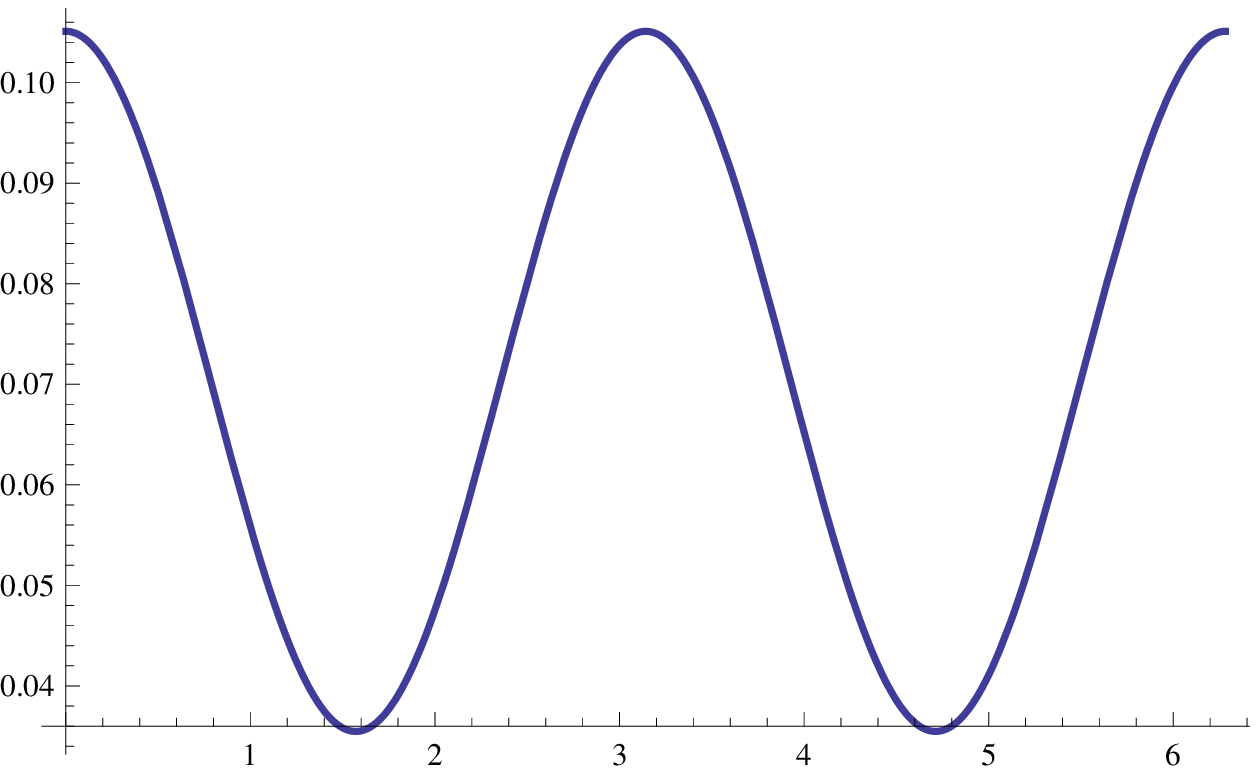}
}\\
{\hspace{-2.0cm} {$\Phi \rightarrow$ }}
 \caption{The  width $\delta x$  expected in directional experiments, relative to that of the standard (non directional experiments), as a function of the angle between the  direction of observation and the sun's direction of motion   for various values of $\Phi$ (a). The $\Phi$ dependence, expected due to the asymmetries of the caustic ring distribution,  is quite small. We also show the $\Phi$ dependence if the direction of observation is made in a plane perpendicular to the sun's direction of motion (b)}. The effect is now small.
 \label{fig:cDirWidth}
  \end{center}
  \end{figure}

	The relevant information for our purposes can be extracted from   table V of Duffy and Sikivie \cite{Sikivie08}, which has improved and updated earlier versions, and for the reader's convenience is summarized in table \ref{tab:caustic}.
One then writes the velocity distribution as
\barr
f(\upsilon_{z})&=&\sum_{n\pm } \eta_{\upsilon^{n\pm}_{\phi}}\delta(\upsilon_{z}-\upsilon^{n\pm}_{\phi}),\nonumber\\
f(\upsilon_{x})&=&\sum_{n\pm } \eta_{\upsilon^{n\pm}_{\rho}}\delta(\upsilon_{x}-\upsilon^{n\pm}_{\rho}),\nonumber\\
f(\upsilon_{y})&=&\sum_{n\pm } \eta_{\upsilon^{n\pm}_{z}}\delta(\upsilon_y-\upsilon^{n\pm}_{z})
\earr
where $\upsilon_{x},\,\upsilon_{y},\,\upsilon_{z}$ are the components of the velocity components in our notation. $\eta_{\upsilon^{n\pm}_{\phi}},\,\eta_{\upsilon^{n\pm}_{\rho}},\,\eta_{\upsilon^{n\pm}_{z}}$
are suitable normalization factors extracted from the corresponding densities of the last columns of table \ref{tab:caustic}.
\begin{table}[htbp]
\begin{center}
\caption{The velocities caustic ring velocities in our vicinity  of our galaxy. The components are given in Sikivie's notation. In our notation for the galactic axes we use $\hat{\phi}\rightarrow\hat{z}$,$\hat{\rho}\rightarrow\hat{x}$,$\hat{z}\rightarrow\hat{y}$}
\label{tab:caustic}
\begin{tabular}{|>{$}c<{$}>{$}c<{$}>{$}c<{$}>{$}c<{$}>{$}c<{$}>{$}c<{$}>{$}c<{$}|}
\hline
n&\upsilon^{n\pm }&\upsilon^{n\pm}_{\phi}&\upsilon^{n\pm}_z&\upsilon^{n\pm}_{\rho}&d_n^+&d_n^-\\
  & \mbox{km/s}& \mbox{km/s}& \mbox{km/s}& \mbox{km/s}&10^{-26}\mbox{gr/cm}^{3}&10^{-26}\mbox{gr/cm}^{3}\\
	 1 & 650 & 140 & \pm635 & 0 & 0.3 & 0.3 \\
 2 & 600 & 250 & \pm540 & 0 & 0.8 & 0.8 \\
 3 & 565 & 380 & \pm420 & 0 & 1.9 & 1.9 \\
 4 & 540 & 440 & \pm310 & 0 & 3.4 & 3.4 \\
 5 & 520 & 505 & 0 & \pm120 & 150. & 15. \\
 6 & 500 & 430 & 0 & \pm260 & 6.0 & 3.1 \\
 7 & 490 & 360 & 0 & \pm330 & 3.9 & 1.2 \\
 8 & 475 & 325 & 0 & \pm350 & 1.9 & 1.0 \\
 9 & 460 & 265 & 0 & \pm375 & 1.4 & 0.7 \\
 10 & 450 & 220 & 0 &\pm390 & 0.9 & 0.9 \\
 11 & 440 & 200 & 0 &\pm390 & 0.8 & 0.8 \\
 12 & 430 & 180 & 0 &\pm390 & 0.7 & 0.7 \\
 13 & 420 & 170 & 0 &\pm390 & 0.6 & 0.6 \\
 14 & 415 & 155 & 0 &\pm385 & 0.6 & 0.6 \\
 15 & 405 & 140 & 0 &\pm380 & 0.5 & 0.5 \\
 16 & 400 & 13 & 0 &\pm375 & 0.5 & 0.5 \\
 17 & 390 & 120 & 0 &\pm370 & 0.5 & 0.5 \\
 18 & 380 & 110 & 0 &\pm365 & 0.4 & 0.4 \\
 19 & 375 & 100 & 0 &\pm360 & 0.4 & 0.4 \\
 20 & 370 & 95 & 0 & \pm355 & 0.4 & 0.4 \\\hline
	\hline
\end{tabular}
	\end{center}
	\end{table}
	With these ingredients we compute the modulated widths, see Fig. \ref{fig:cModWidth}. We observe that the difference between the maximum and the minimum is $0.08$. Note, however, that the maximum is shifted from $\alpha=0$ (June third) $\alpha=0.79 \pi$, i.e. to approximately two weeks later. This is expected due to the asymmetries in the caustic velocity distribution. We also have obtained  the directional dependence of the  width, see Fig. \ref{fig:cDirWidth}. We note, however, that in this case the width depends  on the azymouthal angle $\Phi$. This does not occur in the case of symmetric  velocity distributions so long as the velocity of the Earth is neglected. Here the variation can be sizable for directions of observation lying in a plane perpendicular to the sun's velocity, but small away from it.
 This is another special signature of the caustic ring scenario. Anyway in this case one need not know the sense of direction of the axion.
\section{Discussion}
	In the present work we discussed the time variation of the width of of the axion to photon resonance cavities involved in Axion Dark Matter Searches, by considering a number of popular halo models. We find two important signatures:
	\begin{itemize}
	\item Annual variation due to the motion of the Earth around the sun. We find that  in the relative width, i.e. the width divided by its time average, can attain  significant differences  between the maximum expected in June  and the minimum  expected six months later. This variation is larger than the modulation expected  in ordinary dark matter of WIMPs. It does not depend on the geometry of the cavity or other details of the apparatus. 
	It depends somewhat on the assumed velocity distribution.
	\item The width depends strongly of the angle of direction of observation relative to the sun's direction of motion, even if the sense of direction is not known. This can manifest itself as a characteristic diurnal variation due to the rotation of the Earth around its own axis (see our earlier work \cite{SemerVer15} on how one can translate the directional data into diurnal variation).  Anyway once such a device is operating, data  can be taken  as usual. Only one has to  bin them according the  time they were obtained. If a potentially useful signal is found, a complete analysis can be done according the directionality to firmly establish that the signal is due to the axion.
	\end{itemize}
	In conclusion in this work we have elaborated on two signatures that might aid the analysis of axion dark matter searches. Eventually, if such an observation is made, one may be able to exploit the results obtained here to gain information about the velocity distribution associated with the various halo models. 
	

\end{document}

%% file: begin.tex
\newcommand{\beqa}{\begin{eqnarray}}
\newcommand{\eeqa}{\end{eqnarray}}
\newcommand{\bdm}{\begin{displaymath}}
\newcommand{\edm}{\end{displaymath}}